# Daytime passive radiative cooler using porous alumina

Yang Fu, Jiang Yang, YiShu Su, Wei Du and YunGui Ma*

*State Key Lab of Modern Optical Instrumentation, Centre for Optical and Electromagnetic Research, College of Optical Science and Engineering, Zhejiang University, Hangzhou 310058, China*

We report a daytime passive radiative cooler using chemically fabricated porous anodic aluminum oxide (AAO) membranes. Effective medium theory (EMT) has been applied to analyzing the optical properties of the air-doped porous medium. The composite structure is specifically optimized so that it has a high absorbance (emittance) in the far-infrared atmospheric window and nearly no loss in the solar spectrum. The calculated emittance is well reproduced in the experiment by our AAO sample. The fabricated porous membrane shows a potential cooling power density of 64 W/m$^2$ at ambient (humidity = 75%) under direct sunlight irradiance (AM1.5). Experimentally, the sample is cooled by a 2.6 °C temperature reduction below the ambient air temperature in the sunlight. This performance shows little variance at night. The AAO approach proposed here may provide a promising way to produce low-cost and efficient radiative cooler in large scales for feasible energy conservation.
*Published by AIP Publishing*. http://doi.org/10.1063/x.xxxxxxx

Compared with other refrigeration methods, passive radiative cooling can dissipate heat without consuming electricity and is energetically more favorable. In recent years, it has gained increased research interests in particular for air-conditioning (AC) applications[1-3]. Different from the traditional hot-object cooling that requires the cooler to have large light absorbance or emittance over the whole thermal radiative spectrum[4-5], the AC type radiative cooler needs a selected absorbance spectrum, ideally totally opaque in the far-infrared (IR) atmospheric window (8-13 μm) and transparent in the rest. Then it could effectively loss heat energy into the outer empty space (3 K) meanwhile having minimum absorbance of energy from the surrounding. If the cooler is also fully absorption-free in the solar spectrum, it can work for the daytime under direct sunlight irradiance.

So far, passive radiative cooling systems based on various materials and structures such as bulk media (gypsum, painted metal, etc)[6,7], thin films (SiO, PbSe, ZnS, CdTe, MgO, etc)[8-12], and gases (ammonia, ethylene, ethylene oxide, etc)[13,14] have been explored. However, their cooling performance was usually limited because the intrinsic absorption bands of the ingredients cannot well match the desired IR spectrum window for an AC cooler. To improve this, metamaterials with engineered dispersions, for example resonation particles incorporated polymers, have been explored to achieve the desired selective emittance[15-17]. These designs are mostly configured for the condition without solar irradiance, i.e., at night. For energy saving, a cooling system able to work in the daytime with direct solar irradiance is very attractive and also critical for the usage like AC roofs[1,2,12]. It should be either transparent or highly reflective for the sunlight besides the far-IR absorbance. Two basic configurations have been generally developed to meet the selective spectral requirement. The first category is to stack an IR- transparent solar reflector above the IR absorber (emitter), as usually implemented by multilayer coatings or films[18-21]. The second approach is using artificial optical structures (e.g., photonic crystals) to obtain the desired selective absorbance[22-24]. For practice, both methods may have the cost and efficiency issues. The multilayer or artificial structures will necessarily involve expensive vacuum deposition systems and/or advanced nano fabrication instruments. For mass production, a both technically and economically efficient way is highly desired.

In this paper, we show that chemically obtained AAO porous membrane may be a promising material condidate to produce large size low-cost daytime radiative coolers. Alumina has strong acoustic resonance absorption in the far-IR window but the continuous bulk is not directly suitable for cooling application. Air-hole dilution could help to greatly decrease the dielectric constant and thus improve the impedance match with the surrounding medium (air). Different from the previous work where AAO was designed for cooling hot objects[5], we manufacture AAO membranes with large porosity and very weak solar absorbance and examine the potential for AC applications. The calculations using the measured dielectric parameters indicates our sample could have an ambient cooling power capacity of 64 W/m$^2$. Experimentally, we show a 2.6 °C temperature cooling below the ambient by a 21mm-diametered AAO disk over the full day measurement.

Fig. 1(a) gives the sketch of our AAO membrane radiative cooler placed on the aluminum substrate. The solar irradiance will partially penetrate through the porous Al$_2$O$_3$ layer with negligible attenuation and totally reflected back by the bottom metal ground. The Al$_2$O$_3$ layer guarantees that only IR light within the atmospheric window spectrum will be absorbed/emitted, mimicking the blackbody's behavior in the atmospheric window. Like many other polar materials, alumina has phononic resonance optical absorptions at the far-IR window as revealed by the permittivity spectra in the top panel of Fig. 1(b)[25]. The huge coefficient numbers at the phononic resonance could be substantially suppressed by diluting the sample with air holes using chemical anodization[26,27]. Note the second loss peak is out of the primary far-IR atmosphere widow and practically undesired. But its influence on the overall performance is marginal because both Planck's thermal photon density and energy levels are

---

Electronic mail: yungui@zju.edu.cn

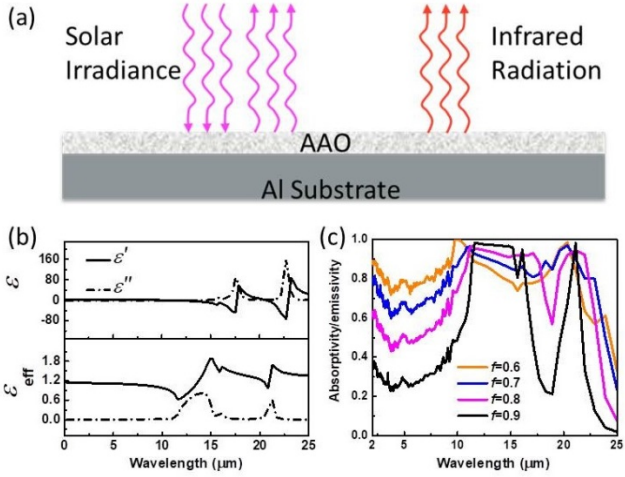

FIG 1. (a) Sketch of an AAO-based radiative cooler. (b) Real part ($\varepsilon'$) and image part ($\varepsilon''$) of dielectric constant for $Al_2O_3$ (top) and AAO membrane at the filling factor of 0.9 (bottom), respectively. (c) Calculated emissivity of AAO membrane at different filling factors.

comparatively low at that region when the device works at room temperature. The size of the etched pores and their unit sizes are usually in the scale of hundreds of nanometers[27]. The oxidized alumina is primarily amorphous and can be characterized by an isotropic permittivity constant. Effective medium theory (EMT) is applied here to evaluating the dielectric constant of the membrane composite for the whole IR region. The one-dimensional anti-hole array will show a uniaxial dielectric property and the effective permittivity could be evaluated using the Bruggeman theory[28]:

$$f \frac{\varepsilon_{\text{air}} - \varepsilon_{\text{eff}}}{\varepsilon_{\text{eff}} + p(\varepsilon_{\text{air}} - \varepsilon_{\text{eff}})} + (1-f) \frac{\varepsilon_{\text{ALO}} - \varepsilon_{\text{eff}}}{\varepsilon_{\text{eff}} + p(\varepsilon_{\text{ALO}} - \varepsilon_{\text{eff}})} = 0 \quad (1)$$

where $\varepsilon_{\text{ALO}}$, $\varepsilon_{\text{air}}$ and $\varepsilon_{\text{eff}}$ represent the dielectric constant of $Al_2O_3$ (obtained from Ref. 25), air and effective medium, respectively. $f$ is the volume filling ratio of air and $p$ is the depolarization factor related to the shape of air inclusions. Here, $p = 1/2$ is used for the in-plane directions.

In the lower panel of Fig. 1(b), we give the in-plane effective permittivity at $f = 0.9$ which is close to our implemented sample (see Figs. 2(a) and 2(b)). From ultraviolet to mid-IR (0.3-20 μm), the real part permittivity is effectively engineered with values less than 1.9 and the image part is visible only at the far-IR phononic resonance bands. Thus, the impedance match with the surrounding air is improved by hole doping. Note the homogenization method is not accurate for the visible and ultraviolet spectra where the operation wavelength is comparable with the hole size and lattice period. According to the Kirchhoff's law of thermal radiation, which explains an object's emissivity is spectrally equal to its absorptivity, Fig. 1(c) gives the emissivity spectrum of the 56.8 μm thick AAO membrane at different hole filling factors calculated from the normal absorbance. At $f = 0.9$, the effective medium has a nearly blackbody-like unit emissivity from 11.6 to 15.3 μm. Reducing the filling factor will decrease the emissivity peak and broaden the bandwidth, which is unwanted for a daytime cooler. Large $f$ will correspond to a better cooling performance. A tradeoff has to been found for practice when mechanical or thermal properties are considered together.

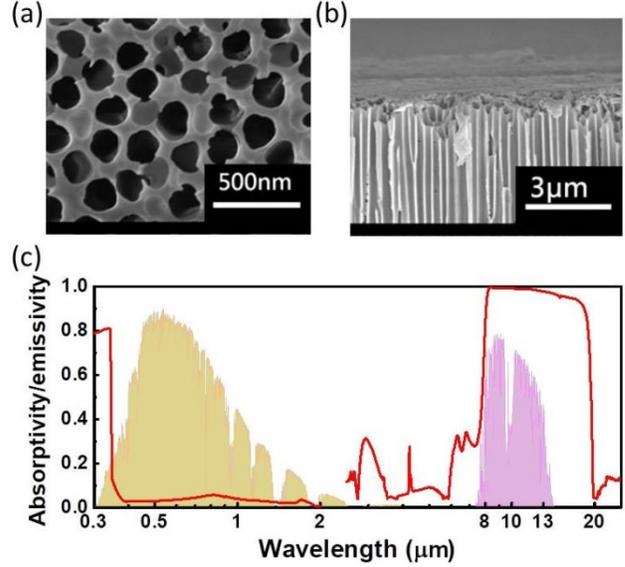

FIG 2. SEM images for (a) top view and (b) section view of AAO sample. (c) Solar spectrum (dark yellow region), atmospheric window at a relative humidity of 70% (light magenta region), measured spectral absorptivity (red line) and emissivity (blue line) of AAO sample.

The standard two-step anodizing method has been employed to fabricate smooth AAO membranes using oxalic acid that helps to acquire a large hole filling factor. After the anodization, the sample is first placed in the $CuCl_2$ solution to remove the rest of Al foil and then transferred into a 5% $H_3PO_4$ solution at 30 °C for 30 min to enlarge the pore size of the membrane. Solution temperature, PH value and anodizing current density are the major controlling parameters for the porous morphology of the sample and the details could be found in previous publications[27]. Figs. 2(a) and 2(b) give the surface and cross-sectional scanning electron microscope (SEM) images of our AAO membrane, respectively. The pore size is estimated to be about 280 nm with a quasi-hexagonal lattice constant about 340 nm, corresponding to a porosity about 0.82. The membrane is 56.8 μm thick and most of the holes are single channel. The rest of aluminum foil after anodization is removed in order to avoid undesired photonic resonance absorption that may be caused by the periodically perturbated dielectric-metal interface as found in the previous experiment[5]. For measurement, the fabricated AAO membrane is adhered to a smooth Al substrate using alumina glue. The reflectivity $R$ of the sample is first characterized by a Fourier transform infrared spectrometer (FTIR) (Bruker Hyperion 1000, Germany) for 2.5-25 μm and a grating spectrometer (Zolix LSH-D30T75, China) for 0.3-2 μm. The latter is attached with an integrating sphere, which is necessary for the measurement of short wavelengths comparable with the lattice period of holes. For the IR measurement, it is not necessary because the side scattering from the high-order diffraction is negligibly small at these long wavelengths for our sub-micron holed samples. Then, the absorptivity (emissivity) is obtained by 1-$R$. As shown in Fig. 2(c) for normal incidence, the measured absorptivity (emissivity) spectrum has a plateau with values larger than 0.9 in 8-20 μm and less than 0.06 in the visible and near-IR regions. Compared with the numerical result (black line) at the closest filling ratio of $f$ = 0.9 shown in Fig. 1(c), the measured far-IR absorption window is broadened and the mid-IR absorptivity is favorably smaller. The quick rise of absorptivity value below 390 nm is not actually real as caused by the first order lattice diffraction; the large diffraction angle is out of the working range of our integrating sphere. In practice, the reflective diffraction is positive by reducing the solar absorbance. The small peak near 3 μm is from the vapor absorption. The ultralow absorptivity in the solar spectrum is very crucial for the daytime cooling usage. Estimated from the measured spectrum, the AAO membrane will have a weak total solar absorbance of 0.054 and a strong far-IR window absorbance of 0.98. The strong absorption at wavelength longer than 14 μm will be negative for an AC cooler.

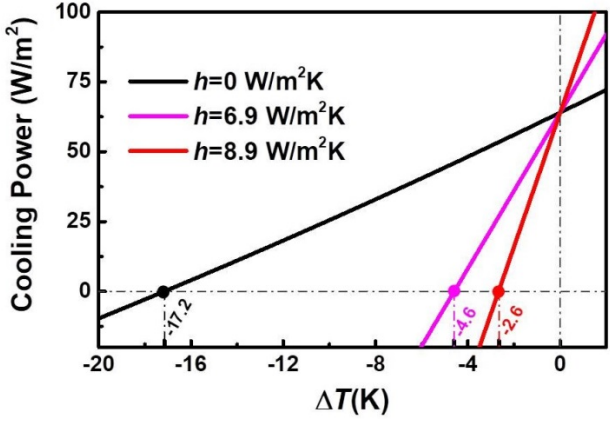

FIG 3. Theoretical cooling power of the membrane sample as a function of temperature difference $\Delta T$ with a nonradiative coefficient $h = 0$ (black), 6.9 W/m$^2$K (magenta) and 18 W/m$^2$K (red), respectively.

In the following, we will explore the practical potential of the fabricated AAO membrane for radiative cooling. Consider a realistic situation, the net cooling power $P_{net}$ per unit radiative surface area at temperature $T$ with the ambient air temperature $T_{amb}$ can be calculated by[24]:

$$P_{net}(T) = P_{rad}(T) - P_{atm}(T_{amb}) - P_{sun} - P_{nonrad} \quad (2)$$

where the power $P_{rad}$ radiated by the cooler is:

$$P_{rad}(T) = \int d\Omega \int_0^\infty e(\lambda, \theta)\cos\theta I_B(T, \lambda)d\lambda \quad (3)$$

with $\int d\Omega = 2\pi \int_0^{\pi/2} d\theta \sin\theta$ is the angular integral over a hemisphere. $I_B(T, \lambda) = \frac{2hc^2}{\lambda^5} \frac{1}{\exp(hc/\lambda k_B T)-1}$ denotes the blackbody radiance with $h$, $k_B$, $c$ and $\lambda$ being Planck's constant, Boltzmann constant, the speed of light in vacuum and wavelength, respectively. The spectral and angular emissivity $e(\lambda, \theta)$ is taken from the measurement. The absorbed power is:

$$P_{atm}(T_{amb}) = \int d\Omega \int_0^\infty I_B(T_{amb}, \lambda)e(\lambda, \theta)\cos\theta e_{atm}(\lambda, \theta)d\lambda \quad (4)$$

from the atmosphere background and:

$$P_{sun} = \int_0^\infty e(\lambda, \theta_{sun})\cos\theta_{sun} I_{AM1.5}(\lambda)d\lambda \quad (5)$$

from the solar irradiance. The spectral angular emissivity of atmosphere is $e_{atm}(\lambda, \theta) = 1 - t(\lambda)^{1/\cos\theta}$ [8], where $t(\lambda)$ is the atmospheric transmittance in the zenith direction[29]. $I_{AM1.5}(\lambda)$ represents the solar irradiance of AM1.5 and $\theta_{sun}$ is the incident angle of sunlight. The last term in Equation (2) represents the nonradiative heat exchange with the surrounding including both

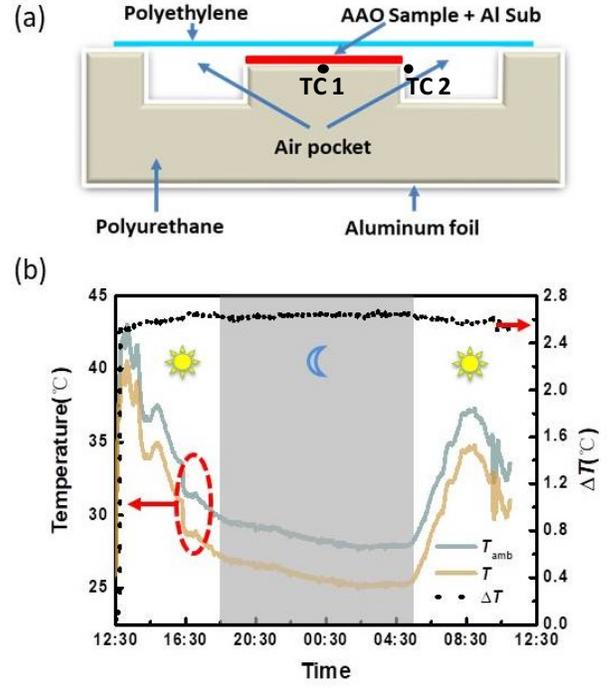

FIG 4. Measurement of the AAO cooler. (a) Schematic of radiative cooling system. (b) Measured 24h temperature for the sample (orange), ambient (dark cyan) and their difference (black dot).

conduction and convection, which could be simply described by:

$$P_{nonrad} = h(T_{amb} - T) \quad (6)$$

where $h$ is the combined nonradiative coefficient which is a key factor affecting the cooling efficiency.

To evaluate the net cooling power with the measured spectral emissivity, we use the atmospheric transmittance data $t(\lambda)$ and the solar irradiance of AM1.5 from the previous publications by incorporating the local relative humidity (~70%)[29,30]. The sunlight incident angle at $\theta_{sun} = 70°$ is used considering the experiment location (Hangzhou city) in the summer. The ambient temperature is 30 °C. Fig. 3 plots the net cooling power changing as a function of the temperature difference $\Delta T (= T - T_{amb})$ between the sample and the ambient air at different $h$. First we see at $\Delta T = 0$, the ambient-temperature sample has a very high net cooling power of 64 W/m$^2$ attributed to the good spectral emittance selectivity of the membrane, which is better than the previous results using photonic structures[8,24,31]. This value will quickly grow to 97 W/m$^2$ in a dry ambient. The cooler's temperature will reduce below the ambient air temperature as the radiated power is much larger than the absorbed. For a vacuum system with $h = 0$, the calculated $P_{net}$ line (black) will lead to a maximum temperature cooling reduction of -17.2 °C using our AAO membrane at $P_{net} = 0$, i.e., at the thermal equilibrium state. The cooling performance is largely influenced by the existence of nonradiative heat transfer. As shown in Fig. 3, the temperature is cooled by -4.6 °C at $h = 6.9$ W/m$^2$ K (magenta) for a typical nonradiative coefficient (mostly with low humidity) often reported in the literature[21,24,32] and -2.6 °C at $h = 8.9$ W/m$^2$ K (red) for a case with high humidity as we meet. The real cooling performance is highly dependent on the environment. The above calculation has assumed a 57 μm thick AAO membrane. Theoretically, the thickness could be controlled as another possible measure to further improve the net cooling power, for example, minimizing the undesired emissivity at the second photonic resonance (around 22 μm). The calculation (not given here) shows this emissivity peak can be suppressed smaller than 0.2 for a 4 μm thick AAO membrane, while the first peak is only slightly narrowed.

On-site cooling experiment is carried out to examine the practical potential of the AAO membrane. The sample is a round disk of diameter of 21 mm glued to an Al sheet. As schematically shown in Fig. 4(a), they are placed inside a thermal box: the side and bottom walls are made of Al foil covered polyurethane (PU) foams and the top is covered by a 50 μm thick polyethylene (PE) film window. The box will help to suppress the nonradiative heat exchange with the external air. Two k-type thermocouples (TCs) as labeled in the figure are used to measure the Al substrate and ambient temperatures, respectively. The measurement was conducted on the rooftop in a typical sunny day (humidity = 70%) in late July 2017, Hangzhou. We collected the 24h temperature data and the results are plotted in Fig. 4(b). The cooler reaches the thermal equilibrium state after 20 min exposure to the sky. The temperature cooling efficiency is about 2.6 °C at night, which is slightly reduced at noon with the direct sunlight irradiance. Thus, the cooling performance is identified. As shown in Fig. 3 by the red line, the measured temperature reduction corresponds to a large nonradiative heat exchange coefficients of $h = 8.9 \text{ W/m}^2\text{K}$, which is much larger than those reported in other experiments[24,32]. The difference primarily arises from our specific atmospheric conditions, in particular we have larger humidity that seriously raises nonradiative heat exchange and the atmospheric emissivity $e_{atm}$ as well [33-35]. The transparency of the atmospheric window will reduce very quickly as humidity grows[8,36]. Moreover, the windy and cloudy weather[37] and the angular selectivity[38] of air will also impact the measurement result. Nonetheless, our experiment has clearly implied the practical potential of AAO membranes in the application of radiative cooling.

In conclusion, a selective emitter composed of thin AAO membrane and metal reflector has been proposed for the passive radiative cooler. The optimized membrane structure has a high emissivity over 0.95 in the far-IR atmospheric window and absorbance less than 0.1 in the solar spectrum. It corresponds to a large cooling power of 64 W/m² at the ambient air temperature. Experimentally, we obtain a 2.6 °C temperature cooling in the ambient. These results indicate a low-cost and mass production for passive radiative coolers is highly possible using surface anodized Al foils. For this purpose, the solar absorbance from the dielectric and metal interface needs to be well controlled. For example, AAO membranes with smaller holes (less than 50 nm) and more smooth features can be fabricated using the sulfuric acid anodizing solution[28]. Smaller lattice constant will help to blueshift the corresponding optical resonance and thus move the interface absorption out of the solar spectrum. In the future, the thickness of AAO membrane could be tuned to further enhance the device performance. It also needs to be pointed out that compared with the multilayer coolers, the porous AAO membrane has a low thermal conductivity and does not suit for the cooling of hot objects which is a different application scenario. Technically, it can also satisfy this condition by filling the nanoholes with conductive materials like metals or oxides as explored in material science[39].

The authors are grateful to the partial supports from NSFC (No. 61775195), the National Key Research and Development Program of China (No. 2017YFA0205700) and the NSFC of Zhejiang Province (Nos. LR15F050001& LZ17A040001).

[1] A. Addeo, L. Nicolais, G. Romeo, B. Bartoli, B. Coluzzi and V. Silvestrini, Sol. Energ. **24**, 93 (1980)
[2] T. M. Nilsson, G. A. Niklasson and C. G. Granqvist, Sol. Energ. Mat. Sol. Cells. **28**,175 (1992)
[3] Y. Zhai, Y. Ma, S. N. David, D. Zhao, R. Lou, G. Tan, R. Yang and X. Yin, Science **355**, 1062 (2017)
[4] J. C. Raymond, D. P. Cox and B. W. Smith, Astrophys. J. **204**, 290 (1976)
[5] J. Lee, D. Kim, C.-H. Choi and W. Chung, Nano Energ. **31**, 504 (2017)
[6] J. Khedari, J. Waewsak, S. Thepa and J. Hirunlabh, Ren. Energ. **20**, 183 (2000)
[7] D. Michell and K. Biggs, Appl. Energ. **5**, 263 (1979)
[8] C. Granqvist and A. Hjortsberg, J. Appl. Phys. **52**, 4205 (1981)
[9] C. Granqvist, A. Hjortsberg and T. Eriksson, Thin Sol. Films. **90**, 187 (1982)
[10] K. Dobson, G. Hodes and Y. Mastai, Sol. Energ. Mat. Sol. Cells. **80**, 283 (2003)
[11] M. Benlattar, E. Oualim, M. Harmouchi, A. Mouhsen and A. Belafhal, Opt. Comm. **256**, 10 (2005)
[12] P. Berdahl, Appl. Opt. **23**, 370 (1984)
[13] E. M. Lushiku and C.-G. Granqvist, Appl. Opt. **23**, 1835 (1984)
[14] A. Hjortsberg and C. Granqvist, Appl. Phys. Lett. **39**, 507 (1981)
[15] M. M. Hossain, B. Jia and M. Gu, Adv. Opt. Mat. **3**, 1047 (2015)
[16] A. R. Gentle and G. B. Smith, Nanolett. **10**, 373 (2010)
[17] L. Zhu, A. Raman, K. X. Wang, M. A. Anoma and S. Fan, Optica, **1**, 32 (2014)
[18] T. M. Nilsson and G. A. Niklasson, Sol. Energ. Mat. Sol. Cells **37**, 93 (1995)
[19] A. Addeo, E. Monza, M. Peraldo, B. Bartoli, B. Coluzzi, V. Silvestrini and G. Troise, Il Nuovo Cimento C **1**, 419 (1978)
[20] Z. Huang and X. Ruan, Int. J. Heat Mass Transf. **104**, 890 (2017)
[21] H. Bao, C. Yan, B. Wang, X. Fang, C. Zhao and X. Ruan, Sol. Energ. Mat. Sol. Cells. **168**, 78 (2017)
[22] E. Rephaeli, A. Raman and S. Fan, Nano Lett. **13**, 1457 (2013)
[23] Z. Chen, L. Zhu, A. Raman and S. Fan, Nat. Comm. **7**, 13729 (2016)
[24] A. P. Raman, M. A. Anoma, L. Zhu, E. Rephaeli and S. Fan, Nature **515**, 540 (2014)
[25] M. R. Querry, Optical constants, MISSOURI UNIV, KANSAS CITY, 28 (1985)
[26] X. Wang and G.-R. Han, Microelectron. Eng. **66**, 166(2003)
[27] A. Belwalkar, E. Grasing, W. Van Geertruyden, Z. Huang and W. Misiolek, J. Membrane Sci. **319**, 192 (2008)
[28] D. Bruggeman, Ann. Phys (Leipzig) **24**, 636 (1935)
[29] A. Berk, G. P. Anderson, P. K. Acharya, L. S. Bernstein, L. Muratov, J. Lee, M. Fox, S. M. Adler-Golden, J. H. Chetwynd, and M. L. Hoke, Proc. SPIE. **6233**, 62331F (2006)
[30] Solar irradiance of AM1.5, http://rredc.nrel.gov/solar/spectra/am1.5/
[31] T.S. Eriksson, S.-J. Jiang, and C.G. Grandqvist, Sol. Energ. Mater. **12**, 319 (1985)
[32] J.-l. Kou, Z. Jurado, Z. Chen, S. Fan and A. J. Minnich, ACS Phot. **4**, 626 (2017)
[33] A. Harrison, Sol. Energ. **26**, 243 (1981)
[34] B. Landro and P. McCormick, Int. J. Heat Mass Transf. **23**, 613 (1980)
[35] P. Berdahl and R. Fromberg, Sol. Energ. **29**, 299 (1982)
[36] M. Hossain and M. Gu, Adv. Sci. **3**, 7 (2016)
[37] M. Hu, B. Zhao, J. Li, Y. Wang and G. Pei, Energy **137**, 419 (2017)
[38] G. Smith and A. Gentle, SPIE Nanostruct. Thin Films II **7404**, 19 (2009)
[39] D. Routkevitch, T. Bigioni, M. Moskovits and J. M. Xu, The J. Phys. Chem. **100**, 14037 (1996)